\documentclass[prb,twocolumn,showpacs]{revtex4}

\usepackage{graphicx}
\usepackage{dcolumn}
\usepackage{amsmath}

\begin{document}
\title{Transport, magnetic, thermodynamic and optical properties in Ti-doped Sr$_{2}$RuO$_{4}$}
\author{K.~Pucher}
%\email{Klaus.Pucher@Physik.Uni-Augsburg.DE}
\author{J.~Hemberger}
\author{F.~Mayr}
\author{V.~Fritsch}
\author{A.~Loidl}
\affiliation{Experimentalphysik~V, Elektronische Korrelationen und
Magnetismus, Institut f\"{u}r Physik, Universit\"{a}t Augsburg,
D-86135~Augsburg, Germany}

\author{E.~W.~Scheidt}
\affiliation{Experimentalphysik~III, Institut f\"{u}r Physik,
Universit\"{a}t Augsburg, D-86135~Augsburg, Germany}

\author{S.~Klimm}
\author{R.~Horny}
\author{S.~Horn}
\affiliation{Experimentalphysik~II, Institut f\"{u}r Physik,
Universit\"{a}t Augsburg, D-86135~Augsburg, Germany}

\author{S.~G.~Ebbinghaus}
\author{A.~Reller}
\affiliation{Festk\"{o}rperchemie, Institut f\"{u}r Physik, Universit\"{a}t
Augsburg, D-86135~Augsburg, Germany}

\author{R.~J.~Cava}
\affiliation{Department of Chemistry and Princeton Materials
Institute, Princeton University, Princeton, NJ 08544,USA}

\date{\today}

\begin{abstract}
We report on electrical resistivity, magnetic susceptibility and
magnetization, on heat capacity and optical experiments in single
crystals of \mbox{Sr$_2$Ru$_{1-x}$Ti$_x$O$_4$}. Samples with
\mbox{$x=0.1$} and 0.2 reveal purely semiconducting resistivity
behavior along $c$ and the charge transport is close to
localization within the $ab$-plane. A strong anisotropy in the
magnetic susceptibility appears at temperatures below 100~K.
Moreover magnetic ordering in $c$-direction with a moment of order
\mbox{$0.01~\mu_{\rm{B}}$/f.u.} occurs at low temperatures. On
doping the low-temperature linear term of the heat capacity
becomes significantly reduced and probably is dominated by spin
fluctuations. Finally, the optical conductivity reveals the
anisotropic character of the dc resistance, with the in-plane
conductance roughly following a Drude-type behavior and an
insulating response along $c$.
\end{abstract}

%\doi{}


\maketitle

\section{Introduction}
After the first synthesis and characterization of
\mbox{Sr$_{2}$RuO$_{4}$} (Ref.~\onlinecite{Randall59}) the system
gained considerable interest after reports of superconductivity
below \mbox{$T \approx 1$~K} by Maeno \emph{et al.}
\cite{Maeno94}. The extremely strong suppression of
superconductivity on non-magnetic impurities
\cite{Mackenzie98a,Mao99} gave first hints on unconventional
superconductivity. That triplet pairing might be favored in
\mbox{Sr$_{2}$RuO$_{4}$} was pointed out in early discussions
\cite{Rice95,Sigrist99,Baskaran96}. And indeed, at present there
exists sound experimental evidence that the superconducting order
parameter is of $p$-wave symmetry: NMR Knight shift
\cite{Ishida98}, muon spin rotation \cite{Luke98} and small angle
scattering from the flux lattice \cite{Kealey00} support the idea
that the superconducting state breaks time-reversal symmetry, not
compatible with either $s$-wave or $d$-wave states. Furthermore,
power-law dependencies of the heat capacity, \mbox{$C \propto T
^{2}$} (Ref.~\onlinecite{Nishizaki99,Nishizaki97}), and of the
spin-lattice relaxation rate, \mbox{$1/T_{1} \propto T ^{3}$}
(Ref.~\onlinecite{Ishida00}), are fingerprints of unconventional
superconductivity.

In analogy to \mbox{$^3$He}, one is tempted to assume that
$p$-wave pairing is mediated via ferromagnetic (FM) spin
fluctuations. Indeed, related compounds are dominated by FM
interactions: \mbox{SrRuO$_{3}$} becomes ferromagnetic below 160~K
(Ref.~\onlinecite{Kanbayasi76}) and \mbox{Sr$_{3}$Ru$_{2}$O$_{7}$}
orders ferromagnetically at 100~K under hydrostatic pressure
\cite{Ikeda00}. However, quite astonishingly there is not much
experimental evidence for ferromagnetic spin fluctuations in the
pure compound. Incommensurate Fermi-surface nesting and
antiferromagnetic (AFM) spin fluctuations have been detected by
inelastic neutron scattering \cite{Sidis99}. In addition, strongly
anisotropic spin fluctuations have been observed in
\mbox{$^{17}$O}-NMR experiments \cite{Mukuda98,Mukuda99} with
significant AFM character. However, similar NMR results by Imai
\emph{et al.} \cite{Imai98} were interpreted to result from
orbital dependent ferromagnetic correlations. The fact that the
presence of FM and AFM spin fluctuations yields a strong
competition between $d$- and $p$-wave superconductivity
\cite{Mazin99} or that spin-triplet superconductivity may even
arise from AFM spin fluctuations \cite{Kuwabara00}, has been
pointed out theoretically. Doping experiments, aiming to induce
long-range magnetic order seem to be important to unravel the
question of the importance of FM vs AFM spin fluctuations in
\mbox{Sr$_{2}$RuO$_{4}$}.

Strontium-ruthenate is almost isostructural to the
\mbox{high-$T_c$} parent compound \mbox{La$_{2}$CuO$_{4}$}. The
superconductivity is carried by the \mbox{RuO$_{2}$} layers within
strongly hybridized oxygen $p$ and ruthenium $d$ states. As has
been pointed out, superconductivity in \mbox{Sr$_{2}$RuO$_{4}$} is
extremely sensitive to defect states and it is clear that
substituting \mbox{Ru$^{4+}$} (\mbox{$4d^{4}$}) by nonmagnetic
\mbox{Ti$^{4+}$} (\mbox{$3d^{0}$}) will suppress
superconductivity. However it seems interesting to check the
closeness of the pure system to a magnetically ordered ground
state and the nature of the magnetism that can be induced by
doping. We recall that \mbox{Ca$_{2}$RuO$_{4}$} is an AFM
insulator \cite{Cao97} while \mbox{Sr$_{2}$IrO$_{4}$} is a weakly
ferromagnetic insulator \cite{Cava94,Carter95}. On substituting Ru
for Ir, Ru exhibits its full local \mbox{$S=1$} moment up to a
critical concentration, beyond which the local moment disappears
\cite{Cava94}. Polycrystalline samples of
\mbox{Sr$_2$Ru$_{1-x}$Ti$_x$O$_4$} with \mbox{$0 < x < 1$} have
been synthesized by Oswald \emph{et al.} \cite{Oswald93} and their
reduction behavior and room temperature resistivity have been
studied. With increasing Ti content the samples were found to show
a higher resistivity while the tendency to be reduced decreases.

While this manuscript was in preparation we became aware of
similar experiments by Minakata and Maeno \cite{Minakata01} who
investigated the electrical resistivity and the magnetic
susceptibility for \mbox{Sr$_2$Ru$_{1-x}$Ti$_x$O$_4$} for Ti
concentrations \mbox{$0 < x < 0.25$}. These authors found local
moment formation exhibiting strong Ising anisotropy. A magnetic
moment of \mbox{$0.5~\mu_{\rm{B}}$} per Ti was calculated and the
magnetic order has been characterized as spin-glass like. Here we
present measurements of the electrical resistivity, the magnetic
susceptibility, the heat capacity and the optical conductivity of
single crystalline material doped with 10~\% and 20~\% Ti. Our
results reveal a strong magnetic and electronic anisotropy of the
doped compounds. On increasing Ti concentration the resistivity
increases and the Sommerfeld coefficient significantly decreases.
In addition we find at low temperatures magnetic ordering is
induced, whose nature is not fully understood.

\section{Experimental details}

Single crystals of \mbox{Sr$_2$Ru$_{1-x}$Ti$_x$O$_4$} were grown
by the floating zone melting technique in a \mbox{CSI
FZ-T-10000-H} furnace. The polycrystalline starting materials were
synthesized by conventional solid state reactions from
\mbox{SrCO$_3$}, \mbox{RuO$_2$} and \mbox{TiO$_2$}. To take into
account the evaporation of some \mbox{RuO$_2$} during the crystal
growth, a 10~\% excess of ruthenium oxide was used. Rods of the
polycrystalline compounds with approximately 7~mm diameter and
100~mm length were pressed and sintered at \mbox{1350~$^{\circ}$C}
for 24~h. For the crystal growth experiments power lamps of 1500~W
each were used. The growth was performed in flowing air (1~l/h)
with a growth rate of 5~mm/h. The seed- and feed-rods were counter
rotated at a speed of 35~rpm. The resulting boules consisted of a
large number of crystals. We found that these crystals can easily
be separated by keeping the boules in air for a few days or by
putting them in water for several hours. The single crystalline
samples examined in this work were platelets with typical
dimensions of 1-3~mm in $a$ or $b$ direction and well below 1~mm
in $c$ direction.

For a structural characterization, small pieces of the single
crystals were powdered and were investigated by x-ray
diffractometry. The Ti-doped samples reveal the same body-centered
tetragonal unit cell and space group ($I4/mmm$) as the pure
compounds. The lattice constants are listed in Tab.~\ref{latcon}
and are in good agreement with earlier published values
\cite{Maeno94,Minakata01}. On increasing $x$, the in-plane lattice
constants slightly increase while $c$ reveals a slight decrease.
However, a uncertainty in the Ti concentration of $\pm3~\%$ can
not be ruled out.

\begin{table}[htbp]
\begin{ruledtabular}
\label{latcon}
\caption{Lattice constants for
\mbox{Sr$_2$Ru$_{1-x}$Ti$_x$O$_4$} at room temperature for $x=0.1$
and 0.2.}
\begin{tabular*}{\hsize}{c@{\extracolsep{0ptplus1fil}}c@{\extracolsep{0ptplus1fil}}c@{\extracolsep{0ptplus1fil}}}
Ti concentration&\multicolumn{2}{c@{\extracolsep{0ptplus1fil}}}{lattice constants in \AA}\\
\colrule \rule{0mm}{3.3mm}$x$&$a,b$&$c$\\
\colrule \rule{0mm}{3.3mm}0&3.8704(1)&12.7435(1)\\
0.1&3.8744(1)&12.7163(1)\\
0.2&3.8775(1)&12.7008(2)\\
\end{tabular*}
\end{ruledtabular}
\end{table}

The two principal components $\rho_{ab}$ and $\rho_c$ of the
electrical resistivity tensor were measured using the Montgomery
method \cite{Montgomery71} for temperatures
\mbox{$0.3~\rm{K}<T<300~\rm{K}$}. The specific heat was
investigated with non-commercial setups employing relaxational
methods \cite{Ott85} at low temperatures (\mbox{$T<4$~K}) as well
as quasi adiabatic and ac methods at elevated temperatures. The
magnetic properties were measured employing a superconducting
quantum interference device ({\sc Quantum Design} MPMS) in a
temperature range \mbox{$1.8~\rm{K}<T<400~\rm{K}$} and in fields
up to 50~kOe.

For the measurements of the optical reflectivity we used two
Fourier-transform IR-spectrometers with a full bandwidth of 50 to
8000~cm$^{-1}$ ({\sc Bruker} IF113v) and 500 to 33000~cm$^{-1}$
({\sc Bruker} IFS 66v/S) together with an {\sc Oxford } Opitstat
cryostat. The polarization dependent reflectivity at room
temperature was investigated using a {\sc Bruker} IRscope II
microscope, which offers the possibility to investigate small
fractions of the sample surface in a range well below 0.1~mm$^2$.
All IR-measurements were carried out on cleaved (not polished)
single crystals.

\section{Results and Discussion}

\subsection{dc resistivity}

\begin{figure}[htbp]
\includegraphics[angle=-90,width=86mm,clip]{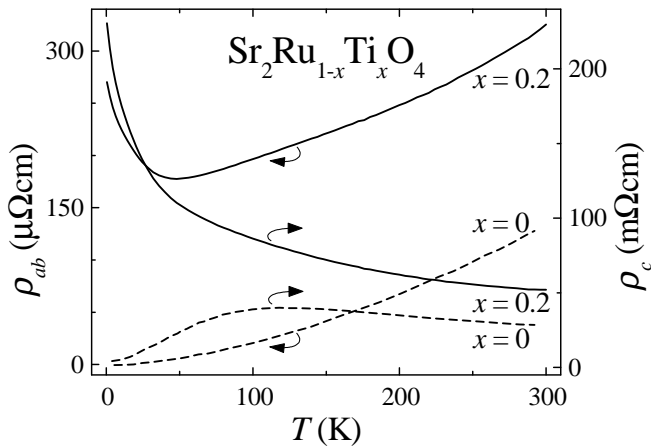}
\caption{Temperature dependence of the in-plane resistivity
$\rho_{ab}$ (left scale) and interplane $\rho_{c}$ (right scale)
in \mbox{Sr$_2$Ru$_{0.8}$Ti$_{0.2}$O$_4$} (solid lines) compared
to the undoped compound (dashed lines) \cite{Lichtenberg92}.}
\label{fig1}
\end{figure}

The interlayer resistivity, $\rho_{c}$, and the in-plane
resistivity, $\rho_{ab}$, were measured for temperatures
\mbox{$0.3~\rm{K}<T<300~\rm{K}$}. As an example Fig.~\ref{fig1}
shows the results for \mbox{Sr$_2$Ru$_{0.8}$Ti$_{0.2}$O$_4$}
(solid lines) compared to the pure compound (dashed lines)
\cite{Maeno94}.  The anisotropy ratio \mbox{$\rho_c / \rho_{ab}$}
increases monotonically as a function of temperature from 160 at
\mbox{$T=300$~K} to 850 at \mbox{$T=0.3$~K}, which are similar to
the ratios of pure \mbox{Sr$_2$RuO$_4$}. At room temperature the
in-plane resistivity (left scale) is enhanced by a factor of 2.5
when compared to the pure compound. On decreasing temperature
$\rho_{ab}$ decreases, passes through a minimum and exhibits a
semiconducting characteristic for \mbox{$T<$~40~K}. This minimum
could signal the onset of localization of charge carriers within
$ab$-plane, Kondo-type scattering of charge carriers on localized
moments or a partial gapping of the Fermi surface due to the
formation of a spin-density wave. We will see later that at 5~K
the optical conductivity reveals a metallic Drude-type of
behavior. In addition we carefully analyzed the resistivity upturn
for \mbox{$x=0.2$} and \mbox{$T<$~50~K} in terms of a Kondo-like
increase or hopping conductivity of localized charge carriers.
Both models do not provide a reasonable description of the
low-temperature upturn. Guided by these facts we prefer to
interpret $\rho(T)$ for low temperatures by the onset of
short-range magnetic order. The interlayer resistivity (right
scale) reveals a semiconducting temperature variation for all
temperatures investigated. For \mbox{$T>100$~K} it is enhanced by
a factor of 2 when compared to the undoped compound, but
$\rho_{c}$ never enters a metallic regime, which is observed for
\mbox{$x=0$} at low temperatures.

\subsection{Magnetic susceptibility and magnetization}

\begin{figure}[htbp]
\includegraphics[angle=-90,width=86mm,clip]{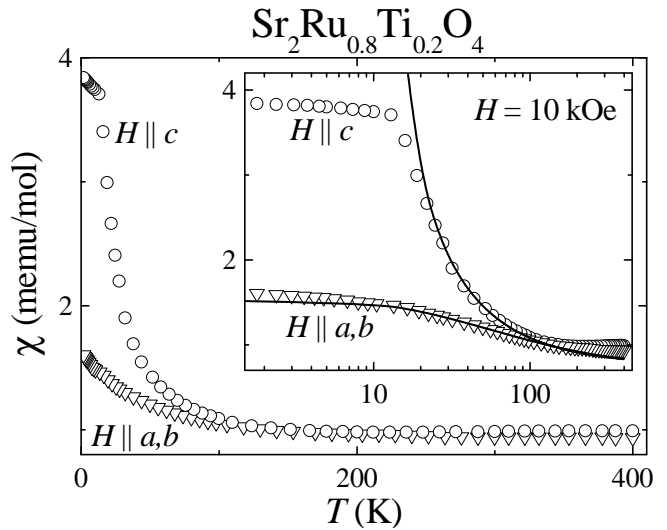}
\caption{Temperature dependence of the dc susceptibility in
\mbox{Sr$_2$Ru$_{0.8}$Ti$_{0.2}$O$_4$} measured with an applied
field of 10~kOe parallel to the $c$-axis ($\chi_{c}$ circles) and
within the  $ab$-plane ($\chi_{ab}$ triangles). The inset shows
$\chi_{ab}(T)$ and $\chi_{c}(T)$ in a semi-logarithmic
plot.}
\label{fig2}
\end{figure}

As a representative result Fig.~\ref{fig2} shows the
susceptibility for \mbox{Sr$_2$Ru$_{0.8}$Ti$_{0.2}$O$_4$} as
measured for an external dc field of 10~kOe. For \mbox{$T>100$~K}
we find an almost isotropic Pauli-spin susceptibility of
approximately \mbox{$10^{-3}$~emu/mol}, very similar to the
results obtained in undoped \mbox{Sr$_2$RuO$_4$}
(Ref.~\onlinecite{Minakata01,Maeno97}). This behavior demonstrates
that at elevated temperatures \mbox{Ti$^{4+}$} \mbox{($3d^{0}$)}
replaces \mbox{Ru$^{4+}$} \mbox{($4d^{4}$)} resulting in
Fermi-liquid (FL) behavior without localized moments. From the
temperature dependence of the resistivity (Fig.~\ref{fig1}) it
seems clear that the FL has quasi two-dimensional (2D) character
as expected from the crystallographic structure. However, below
100~K a Curie-Weiss like behavior evolves and concomitantly a
strong magnetic anisotropy appears, with the $c$-axis
susceptibility $\chi_{c}$ strongly enhanced compared to the
in-plane susceptibility $\chi_{ab}$. This is also true for
\mbox{$x=0.1$}. The results look similar to those reported by
Minakata and Maeno \cite{Minakata01}. An apparent Curie-Weiss (CW)
law for \mbox{$T<150$~K} is followed by a nearly temperature
independent isotropic Pauli-like behavior at elevated
temperatures.

We attempted to fit $\chi(T)$ using the sum of a temperature
independent Pauli spin susceptibility $\chi_{\rm{Pauli}}$ and a
CW-contribution
\begin{equation}\label{eq1}
 \chi (T) = \chi_{\rm{Pauli}} + \frac{A}{T-\Theta}
\end{equation}
At elevated temperatures the Pauli spin susceptibility
contribution is enlarged compared to the contribution of the
localized moments. Therefore no clear prediction is possible if
either the localized moments are still existing but hard to detect
or the localized moments disappear. The best fit results are
indicated as solid lines in the inset of Fig.~\ref{fig2}, which
shows $\chi_{c}$ and $\chi_{ab}$ vs $T$ on a semi-logarithmic plot
to demonstrate the quality of the fit. The corresponding values
for the effective paramagnetic moment $\mu_{eff}$, the
CW-temperature $\Theta$, and the Pauli-contribution
$\chi_{\rm{Pauli}}$ are given in Tab.~\ref{fitpara}. Certainly the
fit of Eq.~\ref{eq1} to $\chi(T)$ is not convincing at high
temperatures. This is due to the fact that for \mbox{$T>100$~K}
$\chi(T)$ slightly increases on increasing temperature. It is this
behavior which led Neumeier \emph{et al.} \cite{Neumeier94} to add
a term which is linear in $T$. It is worth to mention that the
data can be well fitted in the complete range up to room
temperature employing this additional linear term $\chi_{cor}T$
without having significant influence on the results for the
parameters $\chi_{\rm{Pauli}}$, $\mu_{eff}$, and $\Theta$.

\begin{table}[htbp]
\caption{Parameters as determined by the fits Eq.~\ref{eq1} to the
magnetic susceptibility of \mbox{Sr$_2$Ru$_{1-x}$Ti$_x$O$_4$}.}
\label{fitpara}
\begin{ruledtabular}
\begin{tabular*}{\hsize}{l@{\extracolsep{0ptplus1fil}}c@{\extracolsep{0ptplus1fil}}c@{\extracolsep{0ptplus1fil}}c@{\extracolsep{0ptplus1fil}}c@{\extracolsep{0ptplus1fil}}}
Ti concentration&\multicolumn{2}{c}{$x=0.1$}&\multicolumn{2}{c}{$x=0.2$}\\
\colrule \rule{0mm}{3.3mm}$H\|$&$a,b$&$c$&$a,b$&$c$\\
\colrule \rule{0mm}{3.3mm}$\chi_{\rm{Pauli}}$ (emu/mol)&$6 \, \times \, 10^{-4}$&$6 \, \times \, 10^{-7}$&$7 \, \times \, 10^{-4}$&$7 \, \times \, 10^{-7}$\\
$\mu_{eff}$ ($\mu_{\rm{B}}$/f.u.)&$0.73$&$0.47$&$0.65$&$0.50$\\
$\Theta$ (K)&$-100$&$2.5$&$-58$&$5.8$\\
\end{tabular*}

\end{ruledtabular}
\end{table}

An alternative interpretation for the deviation from Pauli
behavior at elevated temperatures rests on the assumption that for
\mbox{$T>100$~K} \mbox{Sr$_2$Ru$_{1-x}$Ti$_x$O$_4$} behaves like a
2D antiferromagnet with a large exchange constant. In these
systems the exchange corresponds to a maximum in the
susceptibility which then would be expected at \mbox{$T>400$~K}
and thereby the increasing susceptibility with increasing
temperature can be well described. Similar observations have been
reported for low-doped \mbox{La$_{2-x}$Sr$_{x}$CuO$_4$}
(Ref.~\onlinecite{Johnston89}).

A parameterisation of the data using Eq.~\ref{eq1} gives strong
AFM correlations within the $ab$-plane (\mbox{$\Theta \approx
-100$~K}) and an almost pure Curie behavior along $c$
(\mbox{$\Theta \approx 0$~K}), indicating that the local moments
are almost decoupled perpendicular to the $c$-axis. Taking this
model seriously \mbox{Sr$_2$Ru$_{1-x}$Ti$_x$O$_4$} has to be
characterized as 2D magnet with a strong in-plane coupling.

\begin{figure}[htb]
\includegraphics[angle=-90,width=86mm,clip]{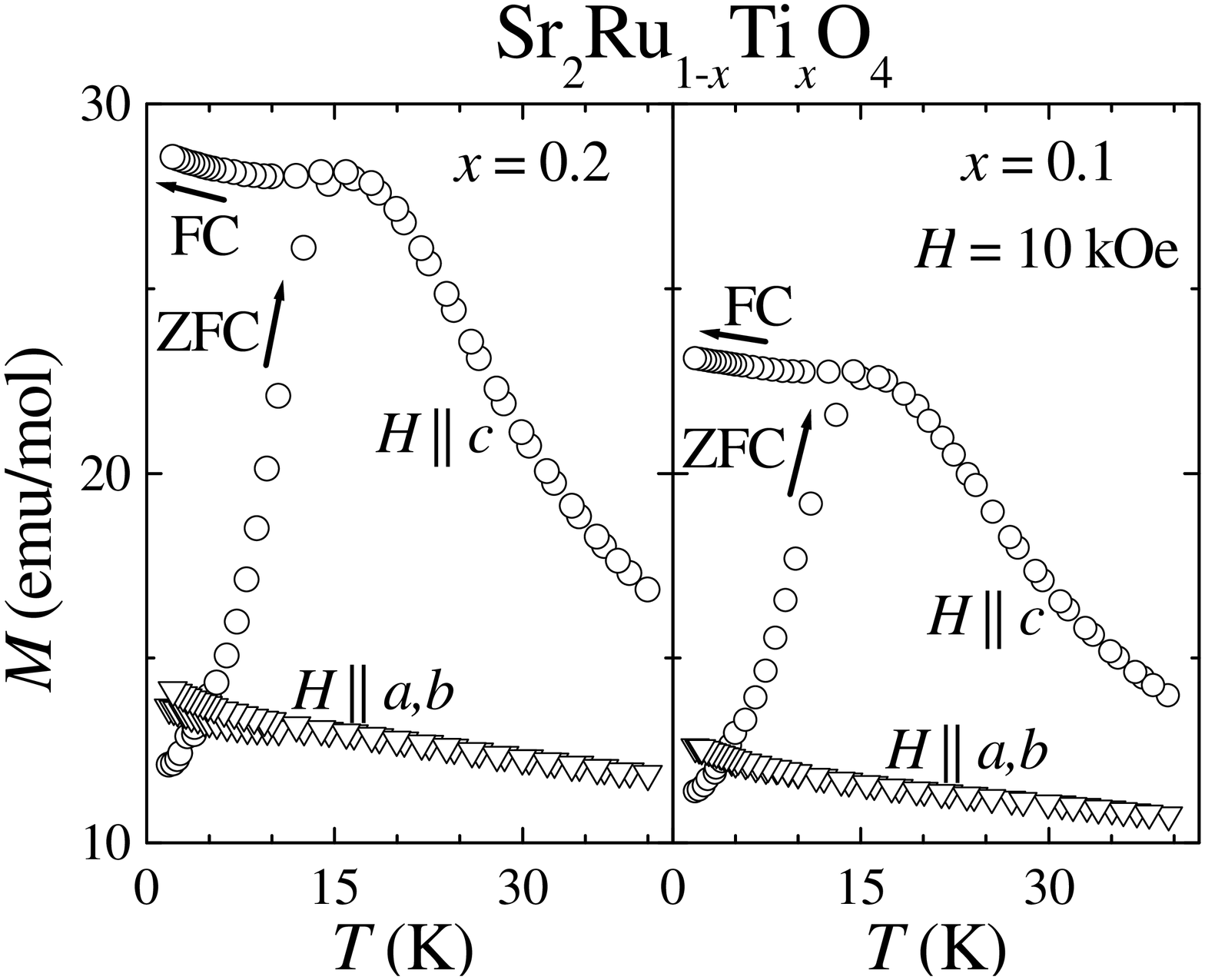}
\caption{Magnetization vs temperature for
\mbox{Sr$_2$Ru$_{1-x}$Ti$_x$O$_4$} for \mbox{$x=0.1$} (right
panel) and \mbox{$x=0.2$} (left panel) for an external field of
10~kOe with \mbox{$H\|c$} and \mbox{$H\|a,b$}. FC and ZFC cycles
are shown.} \label{fig3}
\end{figure}

For the $c$-direction the values of the paramagnetic moment
\mbox{$\mu_{eff} \approx 0.5~\mu_{\rm{B}}$}/f.u.~are larger than
the values found by Minakata and Maeno \cite{Minakata01}. The
paramagnetic moments for the $ab$-plane are enhanced compared to
the $c$-axis. Thus the in-plane magnetic properties of the
Ti-doped compounds seem to reflect the properties of pure
\mbox{Sr$_{2}$RuO$_{4}$}, where values of \mbox{$\mu_{eff} \approx
1~\mu_{\rm{B}}$} and \mbox{$\Theta \approx -150$~K} were reported
\cite{Neumeier94}.

Figure~\ref{fig3} displays the temperature dependence of the
magnetization for lower temperatures for both Ti concentrations. A
clear splitting of the field-cooled (FC) and zero-field-cooled
(ZFC) magnetization occurs close to \mbox{$T_{m}=15$~K} for
\mbox{$H\|c$}. Only minor effects show up for the in-plane
magnetization which might result from a slight misalignment of the
sample. $T_{m}$ is in agreement with the phase diagram published
by Minakata and Maeno \cite{Minakata01}, which shows a saturation
of $T_{m}$ for Ti concentration \mbox{$x > 0.12$}. The fact that
for \mbox{$x=0.1$} and 0.2 the characteristic temperatures $T_{m}$
are almost the same could indicate slight deviations of the
effective Ti concentration from the nominal composition.

One is tempted to assume spin-glass like ordering of statistically
substituted local \mbox{$3d^{1}$} states embedded in a
Fermi-liquid of the band states. However, the FC and ZFC cycles
were performed at relatively high fields of 10~kOe. At such high
fields spin-glass effects often are suppressed \cite{Mydosh93}.
The FC and ZFC splitting can also indicate the formation of FM
clusters or even true long-range ferromagnetism were the FC and
ZFC splitting result from domain-wall effects in strongly
anisotropic materials. Having the time dependence of magnetization
in mind, which has been observed by Minakata and Maeno
\cite{Minakata01}, it seems most plausible to assume short-range
FM correlations only. We also would like to stress that
magnetization for fields within the $ab$-plane is not affected at
all and for low temperatures the zero-field curve with
\mbox{$H\|c$} is well below $M$ with \mbox{$H\|a,b$}. So the
observed magnetic transition is due to a coupling of the moments
along $c$ only.

\begin{figure}[htbp]
\includegraphics[angle=0,width=86mm,clip]{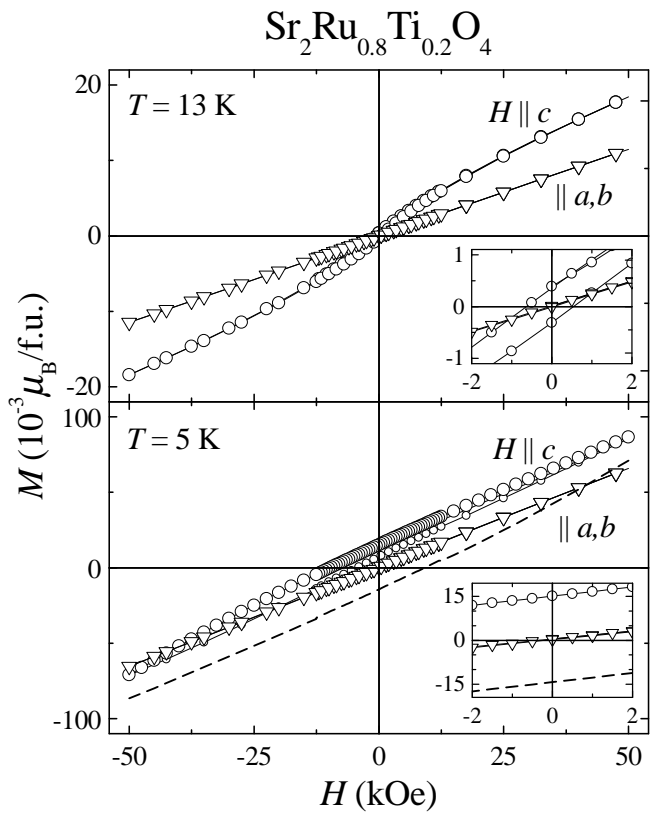}
\caption{Magnetization vs applied field for
\mbox{Sr$_2$Ru$_{0.8}$Ti$_{0.2}$O$_4$} at \mbox{$T=13$~K} (upper
panel) and \mbox{$T=5$~K} (lower panel). Magnetization within the
$ab$-plane  ($M_{ab}$ triangles) and $c$-direction ($M_{c}$
circles) are shown. The dashed line is a mirror image of the
measured $M_{ab}$ data.} \label{fig4}
\end{figure}

Fig.~\ref{fig4} shows the magnetization vs an external magnetic
field below the magnetic ordering temperature (lower frame:
\mbox{$T=5$~K}, upper frame: \mbox{$T=13$~K}). At both
temperatures the in-plane magnetization $M_{ab}$ behaves like a
purely paramagnetic compound (or an AFM well below a spin-flop
field). However, with the applied field along $c$ a clear FM
hysteresis evolves. The coercitive fields rapidly increase towards
lower temperatures. At 5~K the maximum applied field of 50~kOe is
already much too small to establish a complete alignment of the
spins. The ordered moment is rather low, and of the order of
\mbox{$0.01~\mu_{\rm{B}}/\rm{f.u.}$}.

\subsection{Heat capacity}

\begin{figure}[htbp]
\includegraphics[angle=-90,width=86mm,clip]{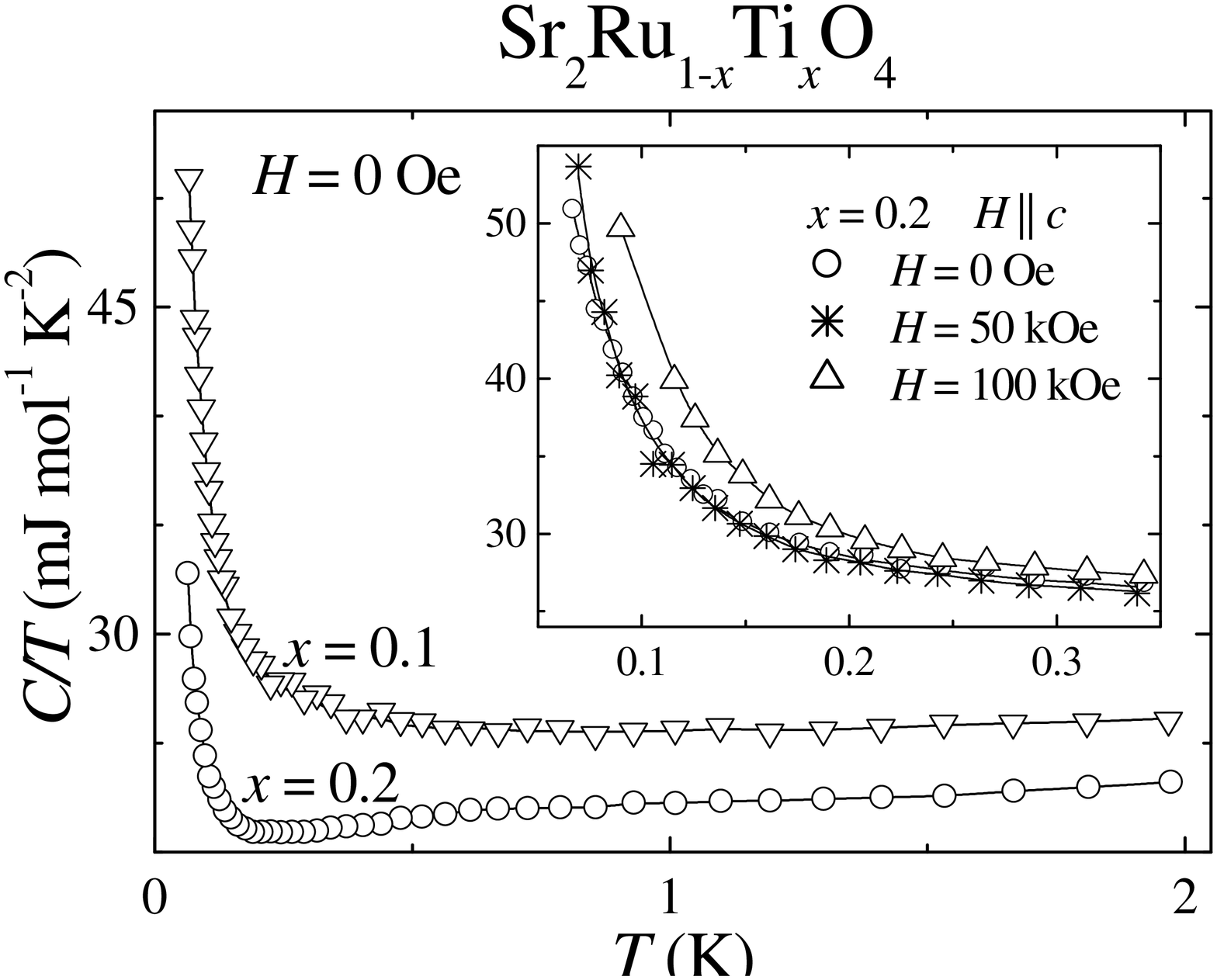}
\caption {Heat capacity $C/T$ vs $T$ in
\mbox{Sr$_2$Ru$_{1-x}$Ti$_x$O$_4$} for \mbox{$x=0.1$} (triangles)
and \mbox{$x=0.2$} (circles) at zero external field. The inset
shows the temperature dependence of \mbox{$C/T$} for external
magnetic fields up to 100~kOe. All solid lines are drawn to guide
the eye.} \label{fig5}
\end{figure}

Figure~\ref{fig5} shows the low-temperature heat capacity $C$ for
\mbox{Sr$_2$Ru$_{1-x}$Ti$_x$O$_4$} for \mbox{$x=0.1$} and 0.2,
plotted as \mbox{$C/T$} vs $T$. With increasing $x$ the
low-temperature plateau, which so far has been interpreted as a
strongly enhanced Sommerfeld coefficient $\gamma$ becomes
suppressed. $\gamma$ is \mbox{40~mJ\,mol$^{-1}$\,K$^{-2}$} in the
pure compound \cite{Nishizaki99,Neumeier94,Nishizaki97} and
becomes reduced to values of approximately
\mbox{27~mJ\,mol$^{-1}$\,K$^{-2}$}
(\mbox{23~mJ\,mol$^{-1}$\,K$^{-2}$}) for \mbox{$x=0.1$} (0.2). A
hyperfine term appears at low temperatures. We have calculated the
hyperfine contributions to the heat capacity resulting from
$^{87}$Sr, $^{99}$Ru and $^{101}$Ru assuming an average magnetic
field and Zeeman splitting. We can fit the low-temperature upturn
assuming a local field of approximately 850~kOe. The local field
seems very strong but still could be reasonable, e.g.~in
\mbox{La$_{0.8}$Sr$_{0.2}$MnO$_3$} local fields about 390~kOe were
detected \cite{Woodfield97}. But these strong internal fields can
be only explained by strong FM correlations and probably point
towards short-range FM order at least. The same fitting procedure
for increasing external fields yields increasing internal fields
and for \mbox{$H=100$~kOe} we found an internal field of
approximately 1250~kOe. The inset in Fig. \ref{fig5} shows the
Schottky-like increase towards low temperatures as a function of
an external magnetic field. Fields significantly higher than
50~kOe are necessary to enhance the hyperfine term. This
observation is compatible with the extremely high saturation
magnetization which is indicated in Fig.~\ref{fig4}.

\begin{figure}[htbp]
\includegraphics[angle=-90,width=86mm,clip]{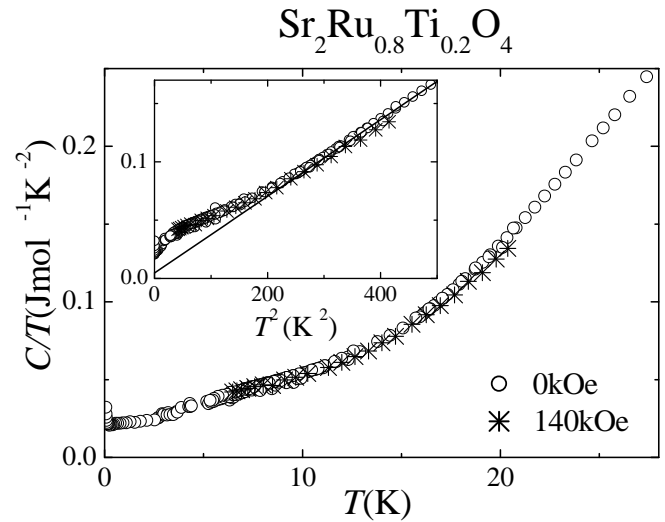}
\caption{Heat capacity \mbox{$C/T$} vs $T$ for
\mbox{Sr$_2$Ru$_{0.8}$Ti$_{0.2}$O$_4$}in zero external field
(circles) and for fields of 140~kOe (crosses). The inset shows the
same data as \mbox{$C/T$} vs $T^2$. The solid line is an
extrapolation of the lattice contribution towards.} \label{fig6}
\end{figure}

Figure \ref{fig6} shows \mbox{$C/T$} vs $T$ for
\mbox{Sr$_2$Ru$_{0.8}$Ti$_{0.2}$O$_4$} for temperatures
\mbox{$0.1~\rm{K}<T<30~\rm{K}$}, in zero external field (circles)
and in a magnetic field of 140~kOe (crosses). \mbox{$C/T$}
smoothly increases with no anomaly, specifically not close to 15~K
where the anomaly in the out-of-plane magnetization has been
detected. For temperatures \mbox{$T>5$~K}, \mbox{$C/T$} reveals no
field dependence up to 140~kOe. In spin-glasses a cusp would be
expected at \mbox{$T\approx1.3~T_g$}, which in our case should
occur close to 20~K. At the onset of long-range magnetic order an
anomaly at \mbox{$T_{m}=15$~K} should show up. Neither anomaly can
be detected in Fig.~\ref{fig6}. The fact that a heat capacity
anomaly is missing favors an interpretation in terms of cluster
ferromagnetism or spin-glass freezing. The inset shows
\mbox{$C/T$} vs $T^2$ and analyzing the heat capacity for
\mbox{$15~\rm{K}<T<25~\rm{K}$} a Sommerfeld coefficient of the
order of \mbox{15~mJ\,mol$^{-1}$\,K$^{-2}$} (solid line) can be
determined. Close to 15~K additional contributions to the heat
capacity show up, which most probably are magnetic in origin. We
would like to mention that a similar analysis could be performed
using the published data for \mbox{Sr$_2$RuO$_4$}
(Ref.~\onlinecite{Neumeier94}) and would result in a much lower
Sommerfeld coefficient. Based on our results we suggest that
\mbox{$C/T$} at low temperatures is due to spin fluctuations even
in the pure compound.

\subsection{Optical conductivity}
We have investigated the reflectivity $R$ of
\mbox{Sr$_2$Ru$_{0.8}$Ti$_{0.2}$O$_4$} in the range of wavenumbers
$\lambda^{-1}$ from 50 to 33000~cm$^{-1}$. Due to the sample
geometry, reflectivity measurements with \mbox{$E\|c$} could only
be preformed using an IR microscope, which operates in the MIR
range ($700~\rm{cm}^{-1}<\lambda^{-1}<7000~\rm{cm}^{-1}$) only.

\begin{figure}[htbp]
\includegraphics[angle=-90,width=86mm,clip]{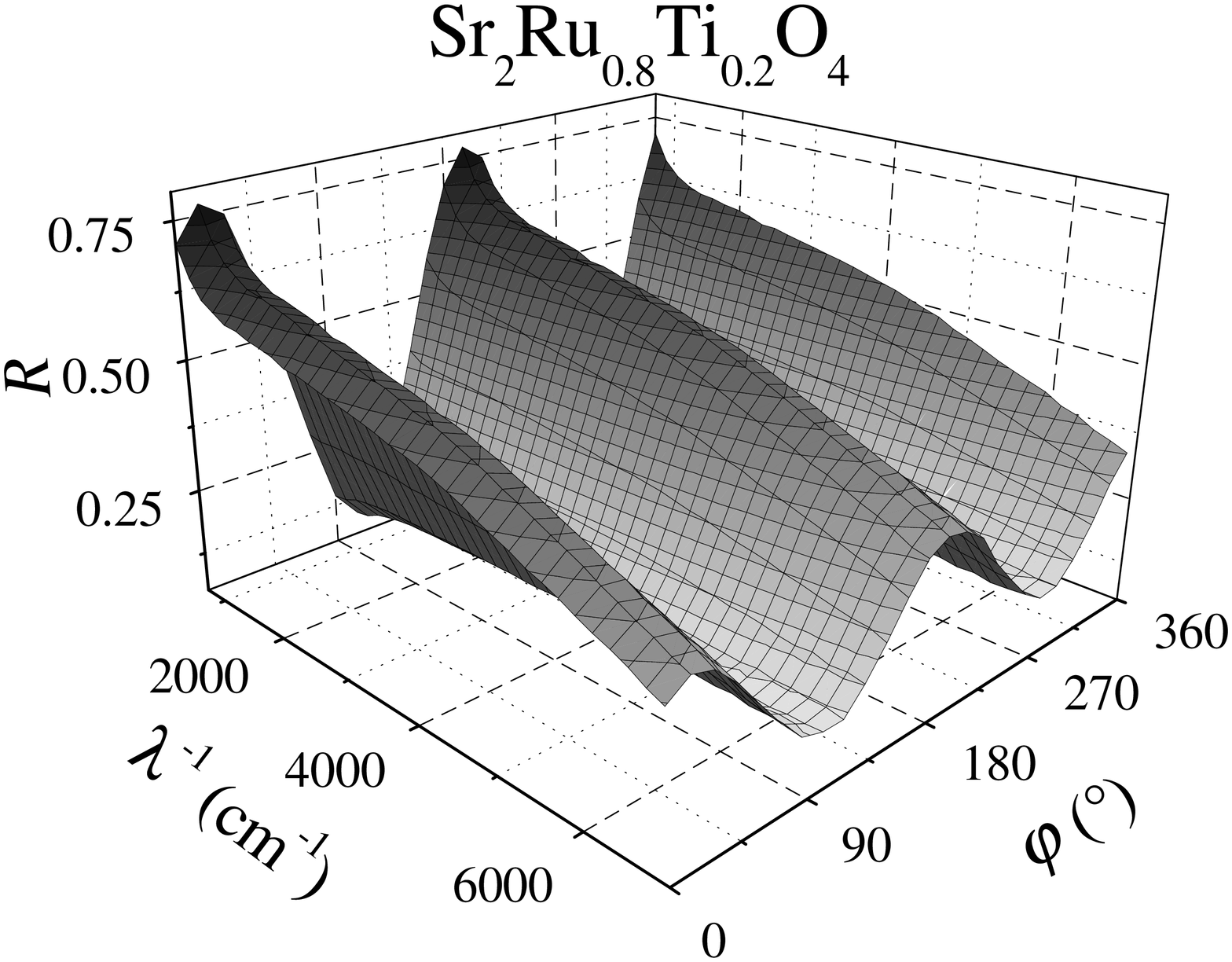}
\caption[]{MIR reflectivity vs wavenumber and direction of $E$
within the $ac$\,-plane for
\mbox{Sr$_2$Ru$_{0.8}$Ti$_{0.2}$O$_4$}.}
\label{fig7}
\end{figure}

The $E$-direction and frequency dependence of the reflectivity at
MIR frequencies for the $ac$-direction of
\mbox{Sr$_2$Ru$_{0.8}$Ti$_{0.2}$O$_4$} is shown in
Fig.~\ref{fig7}. The most striking result is the extreme
anisotropy of the charge dynamics which is nicely documented. The
reflectivity reveals a typical insulating behavior in
$c$-direction (\mbox{$\varphi\approx110^{\circ}$}, $300^{\circ}$)
with nearly constant \mbox{$R\approx0.16$}. In $ab$-direction
(\mbox{$\varphi\approx20^{\circ}$}, $210^{\circ}$) the
reflectivity is much higher (\mbox{$R\approx0.78$} at
700~cm$^{-1}$) and decreases with increasing wavenumber.

\begin{figure}[htb]
\includegraphics[angle=-90,width=86mm,clip]{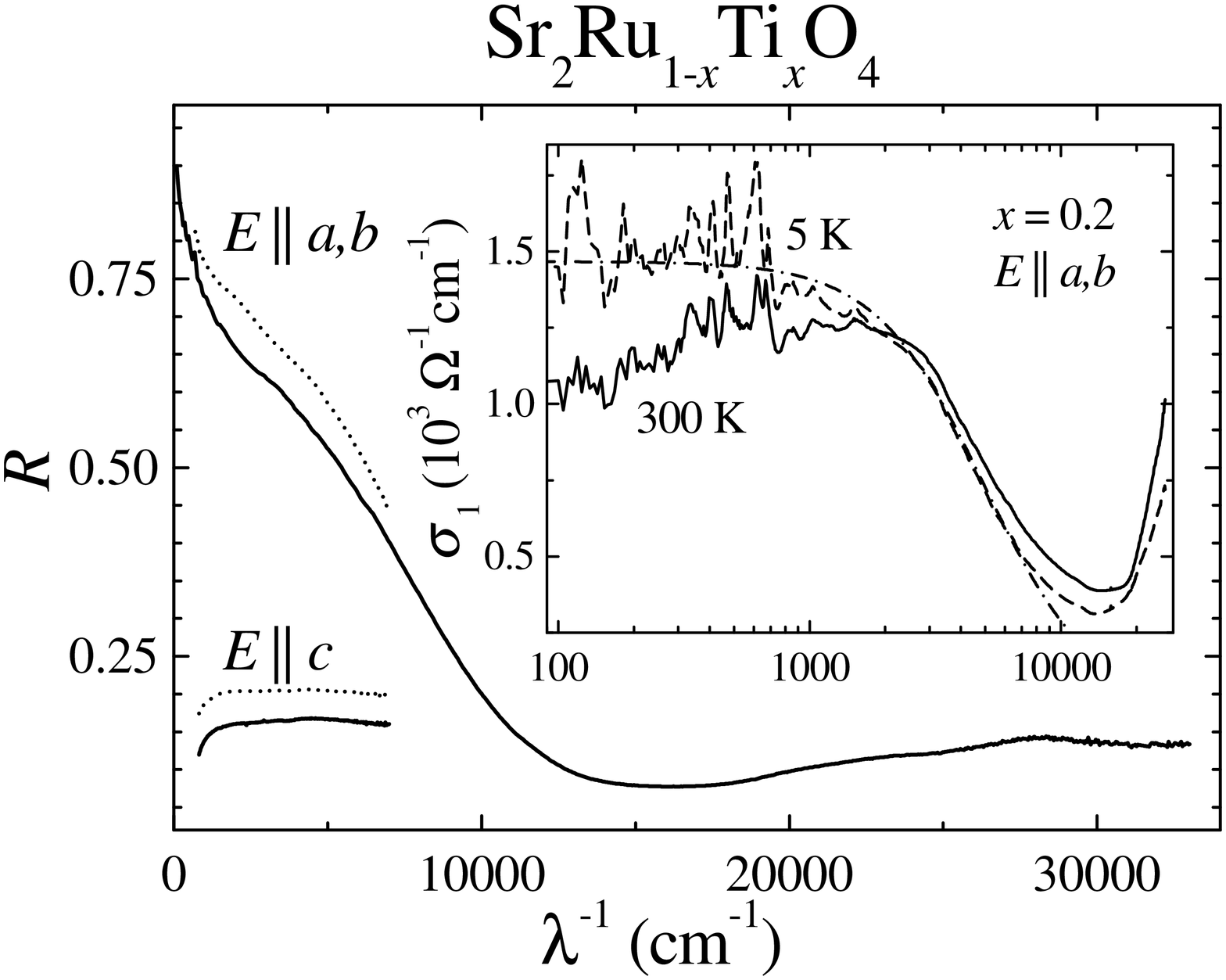}
\caption{In-plane and out-of-plane reflectivity for \mbox{$x=0.2$}
(solid lines) and \mbox{$x=0.1$} (dotted lines) at
\mbox{$T=300$~K}. Inset: Real part of optical conductivity for
\mbox{Sr$_2$Ru$_{0.8}$Ti$_{0.2}$O$_4$} at \mbox{$T=5$~K} (dashed
line), \mbox{$T=300$~K} (solid line) and a Drude fit for
\mbox{$T=5$~K} (dashed-dotted line).}
\label{fig8}
\end{figure}

Figure~\ref{fig8} shows the in-plane reflectivity in a broad
frequency range and the out-of-plane MIR reflectivity for
\mbox{$x=0.2$} (solid lines). The MIR reflectivity for
\mbox{$x=0.1$} (dotted lines) is also shown. The in-plane
reflectivity of \mbox{Sr$_2$Ru$_{0.8}$Ti$_{0.2}$O$_4$} decreases
with increasing temperature in a Drude-like fashion, similar to
observations in pure \mbox{Sr$_2$RuO$_4$}
(Ref.~\onlinecite{Katsufuji96}). The out-of-plane reflectivity
resembles the data of the pure compound, being quite different
from the in-plane $R$. It is nearly frequency-independent at
\mbox{$1500~\rm{cm}^{-1}<\lambda^{-1}<7000~\rm{cm}^{-1}$} and
shows no Drude-like behavior in the measured frequency range. Due
to the sample size for \mbox{$x=0.1$} (dotted lines) only the MIR
reflectivity could be measured. $R$ is enhanced in- and
out-of-plane compared to \mbox{$x=0.2$}, but still is smaller than
the values of the undoped compound \cite{Katsufuji96}. This can be
interpreted as a reduction of free charge carriers with increasing
Ti-doping.

In order to investigate the optical conductivity, a Kramers-Kronig
analysis of the reflectivity was carried out for \mbox{$x=0.2$} at
\mbox{$T=5$~K} and \mbox{$T=300$~K}. For the low frequency
extrapolation the Hagen-Rubens formula has been assumed. The real
part of the optical conductivity \mbox{$\sigma_{1}(\omega)$} is
shown for \mbox{$T=5$~K} (dashed line) and \mbox{$T=300$~K} (solid
line) in the inset of Fig.~\ref{fig8}. The peak-like structures
below \mbox{$1000$~cm$^{-1}$} results from weak reminders of the
phonon peaks and from multiple scattering events of the light
passing the cryostat windows. Turning first to the room
temperature spectra, both the reflectivity and optical
conductivity for \mbox{Sr$_2$Ru$_{0.8}$Ti$_{0.2}$O$_4$} are very
similar to those observed in the related cuprates
\mbox{La$_{2-y}$Sr$_{y}$CuO$_{4}$} (Ref.~\onlinecite{Uchida91}).
In the cuprates the optical spectra for \mbox{$y=0.06$} and
\mbox{$y=0.1$} are very close in the absolute values and shape to
what is observed in the ruthenates under investigation. The slight
non-Drude response towards low frequencies, namely a bump that is
observed close to \mbox{2000~cm$^{-1}$} (see inset in
Fig.~\ref{fig8}) is also a characteristic feature of many
low-doped cuprates. This bump could be disorder induced and
probably is a characteristic feature for metals close to
localization. The fact that the conductivity spectrum for
\mbox{$x=0.2$} at room temperature is close to the spectra
observed in \mbox{La$_{1.9}$Sr$_{0.1}$CuO$_{4}$} is a further hint
for the extremely low charge carrier concentration of the
titanium-doped ruthenates.

To compare the effective number of charge carriers $N_{eff}$ in
the $ab$-plane with the pure \mbox{Sr$_{2}$RuO$_{4}$} we
calculated the spectral weight for the 20~\% Ti-doped sample up to
12000~cm$^{-1}$ (\mbox{$\approx$ 1.5~eV}). $N_{eff}$ is given by

\begin{equation}\label{eq2}
N_{eff}(\omega)=\frac{2\,m_{0}}{Ne^2\pi}
\int_0^\omega\sigma_{1}(\omega')~d\,\omega'
\end{equation}

where $m_{0}$ is the free-electron mass, $e$ is the elementary
charge and $N$ is the number of Ru and Ti atoms per unit volume.
We find \mbox{$N_{eff}^{300\rm{K}}\text{(per f.u.)}=0.38$} at room
temperature, which is clearly reduced when compared to
\mbox{$N_{eff}^{290\rm{K}}=0.53$} as observed in the pure compound
\cite{Katsufuji96}. On the other hand, if $N$ is taken as the
number of Ru atoms per unit volume, we get a value of
\mbox{$N_{eff}^{300\rm{K}}\rm{(Ru)}=0.48$} which is close to that
of pure \mbox{Sr$_2$RuO$_4$}. At 5~K the effective number of
electrons is only slightly reduced to
\mbox{$N_{eff}^{5\rm{K}}\text{(per f.u.)}=0.36$}, however the dc
conductivity at 5~K is clearly enhanced compared to 300~K. A shift
of the spectral weight towards high frequencies often is observed
in highly correlated electron systems. It also is clear that at
5~K the optical conductivity is close to a Drude-like behavior
which contradicts the observation in the dc resistivity where
localization effects were detected at low temperatures.

We fitted the real part of conductivity with a standard
Drude-model in order to estimate the plasma frequency $\omega_{p}$
and the scattering rate $\gamma$ using

\begin{equation}\label{eq3}
\sigma_{1}(\omega)=\frac{\varepsilon_0\,\omega^2_{p}\,\gamma}{\gamma^2+\omega^2}
\end{equation}

where $\varepsilon_0$ is the dielectric constant of vacuum. At
\mbox{$T=5$~K} Eq.~\ref{eq3} provides a good fit to the data
(dashed-dotted line in the inset of Fig.~\ref{fig8}). We find a
plasma frequency \mbox{$\omega_{p}^{5\rm{K}}=20947$~cm$^{-1}$} and
a scattering rate \mbox{$\gamma^{5\rm{K}}=4958$~cm$^{-1}$}. At
room temperature $\sigma_{1}(\omega)$ looks rather similar, with
the scattering rate \mbox{$\gamma^{300\rm{K}}=6465$~cm$^{-1}$} and
the plasma frequency
\mbox{$\omega_{p}^{300\rm{K}}=22197$~cm$^{-1}$} being slightly
increased.
\mbox{$\sigma_{1}^{5\rm{K}}(\omega\rightarrow0)=1475~\Omega^{-1}\,\rm{cm}^{-1}$}
and
\mbox{$\sigma_{1}^{300\rm{K}}(\omega\rightarrow0)=1270~\Omega^{-1}\,\rm{cm}^{-1}$}
are in rough argeement with the dc conductivity shown in
Fig.~\ref{fig1}.

\section{Summary and Conclusions}
In this work we investigated single crystals of
\mbox{Sr$_2$Ru$_{1-x}$Ti$_x$O$_4$} with concentrations of
\mbox{$x=0.1$} and 0.2 which were grown using the floating zone
melting technique. The crystals show an anisotropic behavior of dc
resistivity and infrared reflectivity similar to that observed in
undoped \mbox{Sr$_2$RuO$_4$}, but the temperature dependence of dc
resistivity is rather different. The resistivity along $c$ reveals
a semiconducting behavior down to the lowest temperatures and the
resistivity within the \mbox{RuO$_{2}$}-planes signals the onset
of localization effects close to 50~K. From the magnetic
susceptibility and the magnetization we conclude that the crystals
exhibit an anisotropic Curie-Weiss behavior at temperatures below
100~K. Moreover magnetic ordering along $c$ with a moment of order
\mbox{$0.01~\mu_{\rm{B}}$/f.u.} evolves at \mbox{$T<15$~K}. At the
moment it is unclear if long-range or short-range magnetism is
established below $T_{m}$. However it is clear that strong AFM
exchange dominates the in-plane properties while FM coupling
between the planes establishes FM correlations along $c$.
Antiferromagnetic order within the planes corresponds to spin
arrangements observed in \mbox{La$_2$CuO$_4$} and
\mbox{La$_2$NiO$_4$} and also were detected in
\mbox{Ca$_2$RuO$_4$} (Ref.~\onlinecite{Braden98}). We also
speculate that even at elevated temperatures the Ti-doped
Ruthenates behave like two-dimensional magnets similar to what has
been observed in Sr-doped \mbox{La$_2$CuO$_4$}
(Ref.~\onlinecite{Johnston89}). High-temperature susceptibility
measurements, even in the pure compound are highly warranted. From
the heat capacity experiments we find that the Sommerfeld
coefficient significantly becomes suppressed on Ti-doping reaching
values of \mbox{27~mJ\,mol$^{-1}$\,K$^{-2}$} for \mbox{$x=0.1$}
and \mbox{23~mJ\,mol$^{-1}$\,K$^{-2}$} for \mbox{$x=0.2$}, which
still are extremely high values for a bad metal close to
localization with a low density of charge carriers. Carrying out
the heat capacity in a broader temperature range, we find that
just below the magnetic ordering transition the susceptibility
becomes considerably enhanced and at low temperatures possibly
soft magnetic excitations dominate the heat capacity. However, at
the magnetic ordering temperature, as observed in the
susceptibility experiments, no heat capacity anomalies indicative
for long-range magnetic order were detected. This can be
interpreted as an additional evidence that only short-range order
exists below $T_{m}$. It is interesting to note that the heat
capacity for \mbox{$T>1$~K} remains unchanged in fields as high as
140~kOe.

The reflectivity shows an anisotropic behavior of the charge
dynamics similar to the parent compound. The in- and out-of-plane
reflectivities decrease in the complete frequency range
investigated with increasing Ti-doping. The frequency dependence
of the reflectivity and the conductivity are similar to the low
doped regime of \mbox{La$_{2-y}$Sr$_{y}$CuO$_{4}$}
($y\approx0.06$). We fit the in-plane $\sigma_{1}(\omega)$ using a
standard Drude-model, which describes the experimental results
well at \mbox{$T=5$~K}. At this temperature, the plasma frequency
is approximately 2.3~eV, and the scattering rate is of the order
of 0.6~eV. Both quantities increase on increasing temperature. We
also tried to calculate the optical weight of the Drude response
and found an effective number of charge carriers of 0.38 (per f.u.
at 300~K) assuming that both, Ru and Ti atoms contribute to the
band states. This value is significantly reduced when compared to
the number of charge carriers in the pure system
(\mbox{$N_{eff}=0.53$}) as determined by Katsufuji \emph{et al.}
\cite{Katsufuji96}.

\begin{acknowledgments}
The authors gratefully acknowledge enlightening conversation with
Christoph Langhammer. This work was partly supported by the DFG
via the Sonderforschungsbereich 484 (Augsburg) and project Eb
219/1-1 and partly by the BMBF via the contract
\mbox{EKM/13N6917/0}.
\end{acknowledgments}

\end{document}